\documentclass[%
 reprint,
 amsmath,amssymb,
 aps,
]{revtex4-2}

\usepackage{graphicx}% Include figure files
\usepackage{dcolumn}% Align table columns on decimal point
\usepackage{bm}% bold math
\begin{document}

\preprint{APS/123-QED}

\title{Exceptional Points in Hybrid-Plasmonic Quasiparticles for Ultracompact Modulators}% Force line breaks with \\

\author{Shahab Ramezanpour}
 \email{shahab.ramezanpour@utoronto.ca}
\author{Amr Helmy}%
 \email{a.helmy@utoronto.ca}

\affiliation{
University of Toronto, The Edward S. Rogers Sr. Department of Electrical and Computer Engineering, University of Toronto, 10 King’s College Road, Toronto, Ontario M5S 3G4, Canada
}%

\begin{abstract}
Current progress in electro-optical modulation within silicon integrated photonics, driven by the unique capabilities of advanced functional materials, has led to significant improvements in device performance. However, inherent constraints in dimensionality and tunability still pose challenges for further innovation. In this work, we propose a strategy that exploits the principles of non-Hermitian physics—specifically, the concept of exceptional points (EPs)—to transcend these limitations and pave the way for the next generation of versatile, high-performance photonic devices. Our multilayer structure supports hybrid plasmonic waveguide modes that can manifest as various orders of quasiparticles. By judiciously setting spatial parameters, the system can be tuned to exhibit both weak and strong coupling regimes between the plasmonic and dielectric modes, leading to the controlled formation of EP degeneracies. Furthermore, the integration of low-loss phase-change materials (Sb$_2$S$_3$ and Sb$_2$Se$_3$) enables dynamic electrical tuning, resulting in pronounced modulation of propagation loss and transmission coefficients over sub-micron distances. This superior performance not only sets a new benchmark for device responsivity and compactness but also opens promising avenues for future research, including the incorporation of gain media for loss compensation at EPs and the exploration of alternative tunable materials for next-generation ultracompact photonic devices.
\end{abstract}

\maketitle

%\tableofcontents
\section{Introduction}
A non-Hermitian Hamiltonian is essential for accurately describing physical systems that incorporate non-conservative processes such as gain, loss, and dissipation—effects that traditional Hermitian models, designed for closed, energy-conserving systems, are not well-equipped to capture. This framework leverages fundamental concepts in non Hermitian linear algebra—including parity–time (PT) symmetry—to describe the exceptional point (EP), at which two eigenstates coalesce and the system undergoes a transition from a real to a complex eigenvalue spectrum.
Consequently, these theoretical insights give rise to remarkable effects—such as unidirectional invisibility, enhanced sensitivity, and topological energy transfer—that not only deepen our understanding of wave dynamics but also facilitate the simulation of non-Hermitian behavior in diverse classical systems ranging from photonics and acoustics to mechanical and electronic circuits, by exploiting the formal equivalence between the Schrödinger and classical wave equations\cite{ashida2020non,wiersig2020review,el2018non,li2023exceptional}.

This framework is particularly insightful when extended to light–matter interactions. When light interacts with matter, the collective excitations that arise are not merely the sum of their parts; rather, they manifest as emergent quasiparticles known as polaritons. Unlike fundamental particles, quasiparticles represent collective behaviors resulting from the interplay between light and various excitations in matter such as plasmons, excitons, magnons and phonons \cite{Q1,q2}. Examples include exciton–polaritons in semiconductor microcavities and plasmon–polaritons at metal–dielectric interfaces. These quasiparticles embody hybrid characteristics, merging the properties of photons with those of electronic or vibrational excitations, and play a pivotal role in both fundamental studies and practical applications.

In the realm of integrated photonics, this interplay takes on a new dimension through the formation of waveguide–plasmon polariton quasiparticles. Here, the strong coupling between two distinct modes—the plasmonic mode, which is renowned for its ability to confine light to subwavelength scales, and the waveguide mode, which supports long-range propagation—gives rise to hybrid states with unique optical properties. By carefully tailoring the spatial parameters of the system, it is possible to achieve both weak and strong coupling regimes, each offering distinct advantages. In the strong coupling regime, the hybridization leads to new quasiparticle states that inherit the tight field confinement of plasmonics while maintaining the extended reach of waveguiding. In contrast, the weak coupling regime preserves much of the waveguide’s propagation characteristics while still benefiting from enhanced light–matter interaction \cite{q3,q4,25}.

Crucially, the lossy nature of plasmonic systems means that the formation and dynamics of these waveguide–plasmon polariton quasiparticles are inherently non-Hermitian. The losses associated with the plasmonic component caused by Joule heating, as well as the radiative and material dissipation in the waveguide, require a non-Hermitian description to fully capture the behavior of these hybrid states. This non-Hermitian approach not only provides a deeper understanding of the fundamental physics at play but also opens up new avenues for device engineering, where the sensitivity and tunability of EPs can be exploited to optimize performance.

Exceptional points (EPs) in non-Hermitian systems are singularities where eigenstates and their eigenvalues coalesce, marking a critical transition in the system's behavior. This phenomenon is closely related to parity-time (PT) symmetry, where a PT-symmetric Hamiltonian remains invariant under combined parity and time-reversal operations, yielding real eigenvalues in the balanced PT-symmetric regime. As system parameters vary, an EP emerges at the threshold where PT symmetry is spontaneously broken, transitioning the eigenvalues from being entirely real to forming complex conjugate pairs. Thus, the EP delineates the boundary between the PT-symmetric regime and the broken PT-symmetric regime, a transition that underlies many unique physical phenomena and offers potential for enhanced sensitivity in advanced photonic and sensing applications \cite{1,2,3,4,5,6,7,8,9}. Another unique properties of EP lies in their ability to control the directionality of light in deformed whispering gallery mode (WGM) resonators when operated near a chiral EP \cite{10,11,12,13,ramezanpour2024highly}. Moreover, the interplay between non-Hermiticity and nonlinearity introduces novel functionalities, including the dynamic tuning of EPs \cite{14,ramezanpour2024dynamic,xia2021nonlinear} and the generation of Kerr solitons on both sides of the EP \cite{15}. These advances, along with parallel progress in PT-symmetric systems, have enabled the observation of EP phenomena in a variety of platforms, ranging from hybridized graphene plasmons and vibrational modes \cite{add2} to cavity magnon-polaritons \cite{A9} and both plasmonic \cite{16} and plasmon-excitonic modes \cite{17}, underscoring the broad impact and versatility of EP engineering in modern photonics and condensed matter physics.

Recently our group have revealed that the strong coupling between plasmonic and waveguide modes significantly enhances the sensitivity of Schottky detectors \cite{25}. Nevertheless, the overlapping between the plasmonic and waveguide modes are near 30$\%$. In this work, we demonstrate a near 100$\%$ overlap between plasmonic and waveguide modes—a clear signature of exceptional point (EP) degeneracy—within a meticulously engineered silicon-integrated modulator. Our configuration supports multiple orders of hybrid plasmonic waveguide modes that, through the interplay of plasmonic and dielectric waveguide modes, facilitate both strong and weak coupling regimes, thereby enabling the formation of waveguide–plasmon polariton quasiparticles with EP characteristics. By incorporating low-loss phase-change materials (Sb$_2$S$_3$ and Sb$_2$Se$_3$) as a tunable layer, we achieve remarkable electrical tunability at a wavelength of 1.55 $\mu$m; the hybrid mode exhibits a sharp transition in propagation loss—from approximately 3.14 dB/$\mu$m at the EP to 0.63 dB/$\mu$m out of the EP for Sb$_2$S$_3$, and from 3.00 dB/$\mu$m to 0.89 dB/$\mu$m for Sb$_2$Se$_3$—yielding transmission coefficient variations from around 0.53 (or 0.54) to 0.88 (or 0.83) over a propagation length of just 0.9 $\mu$m. This pronounced sensitivity not only demonstrates a significant step toward advancing both dimensionality and tunability but also sets the stage for future advancements, including gain integration and the exploration of alternative tunable materials, ultimately paving the way for next-generation integrated photonic devices with superior modulation characteristics.
\section{Design and Principles of Operation}
Plasmonics has revolutionized subwavelength optical confinement, offering an effective means to manipulate light at the nanoscale by exploiting surface plasmon polaritons (SPPs). However, despite their ability to localize electromagnetic fields beyond the diffraction limit, the intrinsic dissipative losses in metals present a fundamental challenge, restricting the practical implementation of SPP-based devices in integrated photonics. To mitigate these limitations, hybrid plasmonic waveguides have emerged as a compelling solution, leveraging the interplay between dielectric waveguiding and plasmonic field confinement. By introducing a nanoscale dielectric gap between a dielectric nanowire and a metallic surface, hybrid plasmonic waveguides enable long-range propagation while maintaining strong field confinement, effectively bridging the gap between conventional dielectric waveguides and plasmonic structures \cite{alam,19}.

This hybrid integration has significant implications for silicon photonics, particularly in applications such as microprocessor interconnects and optical sensing. Notably, hybrid plasmonic waveguides have been demonstrated in photodetectors based on asymmetric multilayer configurations, such as $\text{Si-SiO}_2-\text{Al}-\text{Si}$, which exhibit enhanced efficiencies in light-matter interaction, emission, and detection \cite{23}. Moreover, such architectures have facilitated record-high Purcell factors in hybridized plasmonic ring resonators \cite{24,25,26}, underscoring their potential for high-performance optical components. The interplay between plasmonic and waveguide modes, governed by the strength of their coupling, enables the emergence of hybridized quasiparticles known as waveguide-plasmon polaritons. The ability to precisely control the transition between weak and strong coupling regimes further facilitates the realization of tunable non-Hermitian photonic systems, where loss and gain can be leveraged for novel device functionalities, paving the way for enhanced light-matter interactions, ultra-sensitive detection, and dynamically reconfigurable photonic networks.

The Si-Al-Si heterostructure on a SiO$_2$ substrate, where a thin Al layer is embedded between two Si strips, inherently supports two distinct sets of eigenmodes: low-loss waveguide modes confined within the Si layers and high-loss plasmonic modes concentrated at the Si-Al interfaces. Due to the negligible modal overlap, these modes remain spectrally and spatially uncoupled, leading to two independent eigenspaces in the system's Hamiltonian. This separation manifests in the eigenvalue spectrum as modes with significantly different imaginary parts, corresponding to their respective propagation losses.

Introducing a nanoscale SiO$_2$ spacer beneath the Al layer, as in Fig. 1(a), induces a capacitive coupling between the waveguide and plasmonic modes, facilitating both weak and strong modal hybridization depending on the system parameters. This controlled interaction enables the emergence of hybridized waveguide-plasmon polariton modes, where energy exchange between the originally uncoupled modes becomes possible. By precisely tuning the coupling strength through the thickness and width of the layers, the system can be engineered to exhibit hallmark non-Hermitian degeneracies, such as exceptional points (EPs), at which eigenmodes and eigenvalues coalesce.

\begin{figure*}
\centering
{\includegraphics[width=.75\textwidth]{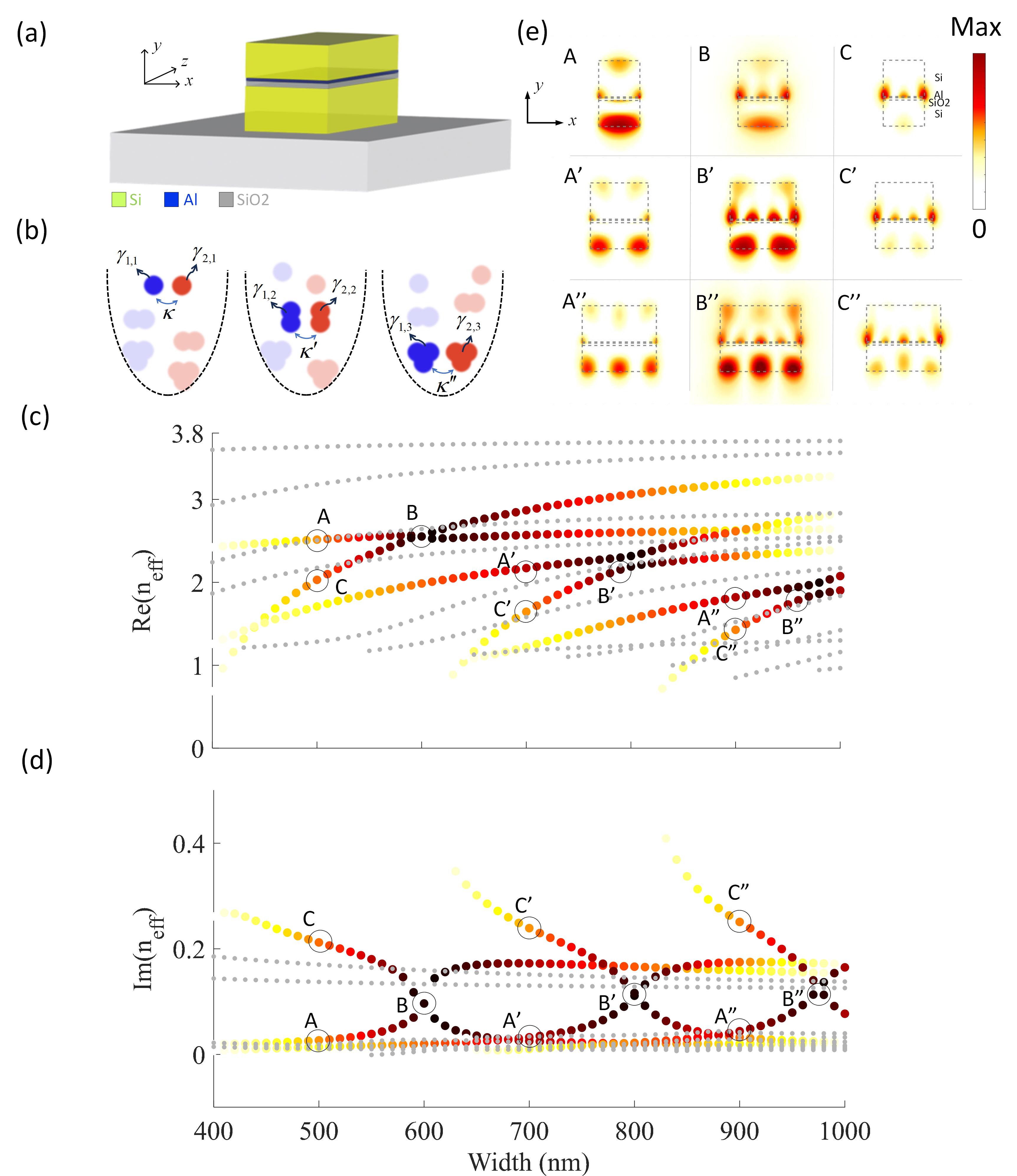}}
\caption{(a)–(b) The layered Si–SiO$_2$–Al–Si structure on an SiO$_2$ substrate, which supports both weak and strong coupling between plasmonic and waveguide modes. This coupling leads to the formation of various orders of waveguide–plasmon polariton quasiparticles that, for specific spatial parameters, exhibit exceptional point (EP) characteristics. (c)-(d) real and imaginary parts of the effective refractive indices, respectively, as functions of the layer width. The highlighted curves correspond to modes that couple strongly to form hybrid plasmonic–waveguide quasiparticles—eventually leading to EP degeneracy—while the grey curves represent uncoupled modes (in the highlighted curves, brighter regions indicate minimal mode overlap; darker regions show increased overlap). Notably, the hybrid plasmonic modes exhibit higher imaginary parts compared to their hybrid waveguide counterparts. (e) For better illustration we depict $E_z$ field distributions at EP and out of EP at selected parameter points: the first column shows field distribution predominantly out of metallic interfaces (characteristic of the waveguide mode), the third column reveals field localization mainly at the metal interfaces (typical of the plasmonic mode), and the middle column corresponds to the nearly degenerate mode at the EP, where the two modes merge.}
\end{figure*}

These hybridized modes, which we refer to as hybridized plasmonic (hybridized waveguide) modes with the acronym HP (HW), however, may primarily have the characteristics of the uncoupled plasmonic (waveguide) mode with higher (lower) imaginary portion of effective refractive index. Exceptional Point (EP), a degeneracy in open systems when both eigenvalues and eigenmodes degenerate, is an interesting feature in non-Hermitian systems. It can be realized by the coupling of either lossy modes or modes with gain and loss. The lossy and (almost) lossless cases in our scenario are HP and HW modes, respectively. We report the first detection of EP degeneracy in hybridized plasmonic and waveguide modes to the best of our knowledge. From the quantum mechanical approach, the structure can be modelled as confined modes within a potential well, in which by judiciary tuning of parameter space supermodes can be generated having EP characteristics.

In the hybrid plasmonic-waveguide system, the transition from weak to strong coupling—and ultimately the tuning of the system to an exceptional point (EP)—is governed by precise control over the modal interaction through structural parameter adjustments. The hybridized plasmonic (HP) and hybridized waveguide (HW) modes, characterized by their complex effective refractive indices, undergo EP degeneracy when two conditions are met: (i) the real parts of their effective indices become identical, and (ii) the difference in their imaginary components equals twice the coupling strength between the modes.

To achieve this delicate balance, we fine-tune the geometric parameters of the system, systematically adjusting the widths and thicknesses of the constituent layers. While modifications to any single parameter inherently influence both the real and imaginary parts of the refractive indices, our numerical simulations indicate that different parameters predominantly affect specific components. Specifically, the width of the layers primarily governs the real part of the effective refractive indices, facilitating spectral alignment of the HP and HW modes. Conversely, the thicknesses of the top and bottom Si layers have a more pronounced effect on the imaginary components, controlling the modal loss disparity and ultimately enabling the fulfillment of the EP condition. For a fixed Al thickness of 10 nm and a SiO$_2$ spacer of 20 nm, the system can be systematically driven from the weak to the strong coupling regime by initially adjusting the layer widths to minimize the real-index mismatch, followed by fine-tuning the Si layer thicknesses to bring the imaginary components into the required balance.

\begin{figure*}
\centering
{\includegraphics[width=.7\textwidth]{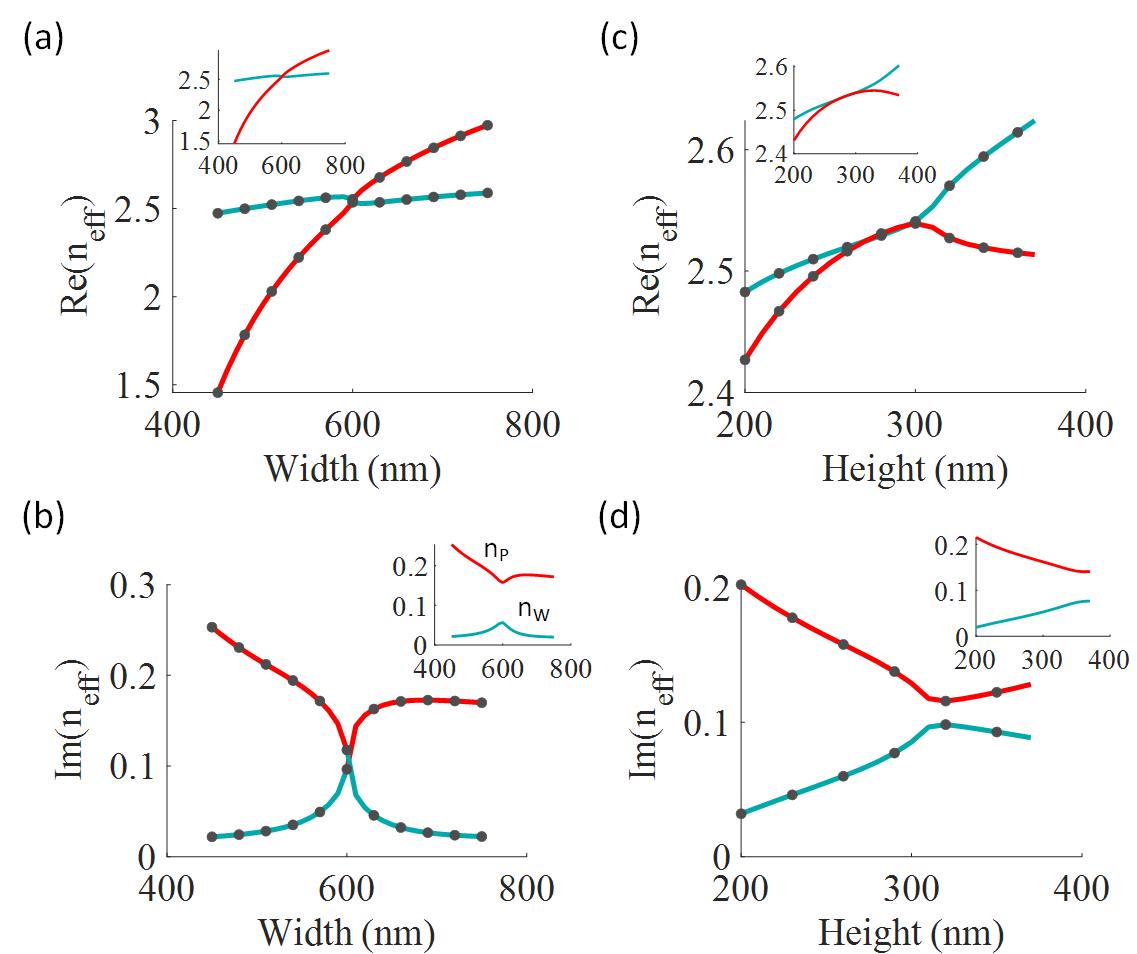}}
\caption{A coupled mode theory analysis is conducted to elucidate the interaction between the waveguide and plasmonic modes, resulting in the formation of hybrid waveguide and hybrid plasmonic modes.  (a)-(b) and (c)-(d) The real and imaginary parts of the effective refractive index as functions of the ridge width and the top silicon height, respectively. The solid curves represent the predictions from coupled mode theory, while the dots denote the mode analysis results obtained from Lumerical simulations. Insets display the equivalent refractive indices of the uncoupled waveguide ($n_W$) and plasmonic ($n_P$) modes extracted via curve fitting, providing additional insight into the modal behavior underlying the hybridization process. At EP, the real part of $n_W$ and $n_P$ are identical while the difference of their imaginary parts are twice of their coupling value. Notably, at the exceptional point (EP), the real parts of $n_W$ and $n_P$ converge, while the difference in their imaginary parts equals twice the coupling value, which result in degeneration of both real and imaginary parts of the effective indices.}
\end{figure*}

Figures 1(c) and 1(d) illustrate the dependence of the real and imaginary parts of the effective refractive indices on the layer widths, for a fixed set of layer thicknesses—20 nm for SiO$_2$, 10 nm for Al, 223 nm for the bottom Si, and 310 nm for the top Si. These plots reveal that, within the hybrid plasmonic waveguide structure, certain hybridized plasmonic (HP) and hybridized waveguide (HW) modes begin to interact, as evidenced by their coupling, while other modes remain uncoupled (depicted in grey). The color variations in the highlighted curves are indicative of the degree of mode overlap. Specifically, the brighter regions correspond to areas where the modes exhibit minimal overlap, while the darker regions indicate increasing overlap between the modes. At maximum overlap—approaching 100$\%$—the modes become degenerate. At a critical point in the parameter space, where both eigenvalues and eigenmodes coalesce, an exceptional point (EP) is formed. This degeneracy is accompanied by abrupt changes in both the real and imaginary components of the effective indices, signifying a transition in the modal behavior. The intermodal coupling between multiple orders of plasmonic and waveguide modes gives rise to a spectrum of hybridized waveguide-plasmon polariton quasiparticles, each exhibiting distinct exceptional point (EP) characteristics, unveiling rich non-Hermitian physics in a multilevel coupling mechanism.

The hybrid waveguide supports multiple plasmonic and dielectric modes, which can be viewed as quasiparticles occupying discrete energy levels in the schematic wells of Fig. 1(b). At a given ridge width, two of these quasiparticles couple strongly (indicated by their brighter shading), forming a hybrid waveguide–plasmon polariton, while the remaining modes remain effectively uncoupled. As the width increases further, successive higher order mode pairs similarly enter into strong coupling, generating a series of higher order waveguide–plasmon polariton quasiparticles.

By fine-tuning the layer dimensions, this structure provides a versatile platform for engineering passive PT-symmetric systems based on hybrid plasmonic-waveguide modes. Furthermore, incorporating active media within the layered configuration may enable the ideal PT-symmetric regime by effectively nullifying the imaginary part of the effective indices at the EP \cite{add5}. In addition, the dielectric layer, traditionally composed of SiO$_2$, can be substituted with a few nanometers of a low-index material or even two-dimensional materials such as graphene \cite{add1,add2,add3} to enhance regulatory performance. The deep subwavelength field confinement in this dielectric region also allows for the integration of ITO, whose epsilon-near-zero (ENZ) properties \cite{add4} can further modulate the device's operation between on- and off-EP regimes. Moreover, the introduction of nonlinear materials offers an additional degree of freedom, enabling dynamic tuning of the structure across various operational phases, thereby paving the way for ultracompact modulation, sensitive detection, and reconfigurable integrated photonic interconnects.
 
The dominant propagation loss in our hybrid structure originates from the overlap of the guided mode with the thin Al film: optical power that resides in close proximity to the metal experiences Ohmic damping, whereas power confined away from the metal remains low‐loss. To visualize this, we plot the amplitude of longitudinal field component $|{{E}_{z}}|$ in Fig. 4(e), which peaks at the metal–dielectric interface and decays into the surrounding dielectrics—this is the signature of a plasmonic mode. Modes exhibiting stronger $|{{E}_{z}}|$ overlap with the Al layer incur higher attenuation. In contrast, the total field amplitude $|E|$ confirms that the low‐loss dielectric‐slot character confines most of the energy within the SiO$_2$ layer between the top and bottom Si slabs (Fig. 4(e)).
As the ridge width is tuned toward the EP, the power distribution of low loss mode—initially concentrated beneath the Al film—gradually shifts upward into the top Si region, while the higher loss plasmonic mode—originally confined above the Al layer in the top Si slab—shifts downward. At the exceptional point, these two modal power distributions become equal, yielding balanced attenuation.

As shown in Figure 1(e), the distribution of \(|E_z|\) for the eigenmodes is presented at several selected points highlighted in Figures 1(c)–1(d). At the degenerate point, where an exceptional point (EP) is realized, the field profiles of the coupled modes converge, exhibiting nearly identical distributions—a clear signature of EP degeneracy. In contrast, away from this degenerate condition, the field distribution diverges: the hybridized plasmonic (HP) mode exhibits predominant localization at the Al interface, while the hybridized waveguide (HW) mode is primarily confined to the bottom Si layer.

The dominant source of propagation loss in our hybrid structure is Ohmic dissipation in the metallic layer, which is determined by the portion of the modal electric field that overlaps the metal — in our geometry this is primarily the $|E_z|$ component. When the slot-like mode is concentrated in the SiO$_2$ layer but has little $|E_z|$ at the metal interface (first column of Fig. 1(e)), the mode is effectively dielectric-slot-like and supports relatively long-range propagation despite the nearby metal. By contrast, when the modal field exhibits strong $|E_z|$ overlap with the Al layer (third column of Fig. 1(e)), the mode becomes plasmonic-like and experiences significantly larger loss. In the presence of both SiO$_2$ and Al, the observed modes are hybridizations of these limiting cases (lossless dielectric slot and lossy plasmonic modes); at certain geometries these hybridized branches approach degeneracy and display the characteristic signatures of operation near an exceptional point (real-part coalescence and vanishingly small loss splitting; second column of Fig. 1(e)). The pure plasmonic and pure dielectric-slot modes can be obtained from the same geometry by removing the SiO$_2$ spacer or the Al layer, respectively.

We note that in realistic, lossy photonic systems such as our hybrid plasmonic waveguide, an ideal exceptional point (EP)—defined by the exact coalescence of both eigenvalues (real and imaginary parts) and eigenvectors of a $2\times2$ non-Hermitian Hamiltonian—cannot be perfectly realized. Physically, the structure inherently supports at least two distinct modal branches (the hybrid-plasmonic and hybrid-waveguide branches), and numerically finite mesh resolution and solver tolerances impose a lower bound on how closely the imaginary parts of the effective indices can match. Nonetheless, we observe all of the characteristic signatures of operation in the vicinity of an EP: real-part coalescence of the eigenvalues, merging of the eigenvectors (spatial field profiles), and vanishingly small loss splitting. To reflect this practical reality and avoid confusion, we therefore use the terms “near EP” and “near 100$\%$
overlap” throughout the manuscript.

In our analysis, the hybrid plasmonic-waveguide system is effectively modeled as an independent two-level system wherein each plasmonic mode couples uniquely to a corresponding waveguide mode, as illustrated in Figures 1(c)–1(d). The spatial evolution of the modal amplitudes, \(a_P\) and \(a_W\), is governed by the \(2\times2\) coupled mode equation
\begin{equation}
\frac{d}{dz}\begin{pmatrix} a_P \\ a_W \end{pmatrix} = -i\begin{pmatrix} k_P & k_0\kappa \\ k_0\kappa^* & k_W \end{pmatrix}\begin{pmatrix} a_P \\ a_W \end{pmatrix},
\end{equation}
where \(k_P\) and \(k_W\) denote the propagation constants of the plasmonic and waveguide modes, respectively, and \(\kappa\) represents the coupling coefficient. By assuming a \(z\)-dependence of the form \((a_P,\, a_W) = (A_P,\, A_W)e^{-ikz}\), the equation reduces to an eigenvalue problem:
\begin{equation}
k\begin{pmatrix} A_P \\ A_W \end{pmatrix} = \begin{pmatrix} k_P & k_0\kappa \\ k_0\kappa^* & k_W \end{pmatrix}\begin{pmatrix} A_P \\ A_W \end{pmatrix}.
\end{equation}

Expressing the propagation constants in terms of effective refractive indices, i.e., \(k = n_{\text{eff}}k_0\), \(k_P = n_P k_0\), and \(k_W = n_W k_0\), we obtain
\begin{equation}
n_{\text{eff}}\begin{pmatrix} A_P \\ A_W \end{pmatrix} = \begin{pmatrix} n_P & \kappa \\ \kappa^* & n_W \end{pmatrix}\begin{pmatrix} A_P \\ A_W \end{pmatrix},
\end{equation}

which leads to the eigenvalue equation
\begin{equation}
n_{\text{eff}} = \frac{n_P + n_W \pm \sqrt{(n_P - n_W)^2 + 4\kappa^2}}{2}.
\end{equation}

, for a real-valued \(\kappa\). An exceptional point (EP) is achieved when the discriminant vanishes:

\begin{equation}
\sqrt{(n_P - n_W)^2 + 4\kappa^2} = 0.
\end{equation}

This condition requires that the real parts of the effective indices become identical,

\begin{equation}
\operatorname{Re}(n_P) = \operatorname{Re}(n_W),
\end{equation}

and that the difference in their imaginary parts satisfies
\begin{equation}
\operatorname{Im}(n_P) - \operatorname{Im}(n_W) = \pm 2\kappa.
\end{equation}

For a typical coupling value of \(\kappa = 0.05\), the appropriate values for \(n_P\) and \(n_W\) are determined to match the results from Lumerical simulations. Figure 2 compares the effective indices obtained from Lumerical (dots) with those predicted by coupled mode theory (solid lines) as a function of the layer width (Figs. 2(a)-2(b)) and the top Si layer height (Figs. 2(c)-2(d)). The inset further displays the extracted values of \(n_{P}\) and \(n_{W}\). The effective indices, $n_P$ and $n_W$, used in the coupled mode theory were determined through curve fitting. Alternatively, these parameters could be obtained by simulating the structure without the low-index material to extract the plasmonic effective index ($n_P$) and without the metallic thin layer to determine the waveguide index ($n_W$). Notably, at the EP, the conditions \(\operatorname{Re}(n_P) = \operatorname{Re}(n_W) = \operatorname{Re}(n_{\text{eff}})\) and \(\operatorname{Im}(n_P) - \operatorname{Im}(n_W) = 2\kappa\) are satisfied, thereby confirming the occurrence of EP degeneracy in our structure and validating the two-level system model through coupled mode theory.

Figure 3 provides a detailed view of the modal evolution near the exceptional point (EP). In the layered structure depicted in Figure 3(a), the device is segmented into six distinct regions (\(S_1\)–\(S_6\)), and the power distribution within these regions is analyzed as a function of the layer width. As shown in Figure 3(c), when operating far from the EP, the hybrid waveguide mode concentrates most of its power in regions \(S_1\), \(S_2\), and \(S_3\); however, as the system approaches the EP, a significant transfer of power occurs toward region \(S_5\). Conversely, for the hybrid plasmonic mode, Figure 3(d) reveals that power is predominantly localized in region \(S_5\) when far from the EP, but it gradually shifts to regions \(S_1\), \(S_2\), and \(S_3\) near the EP. This pronounced power transfer between regions \(S_1\)–\(S_3\) and \(S_5\) is indicative of the balanced field distribution characteristic of EP degeneracy.

To quantitatively assess the coupling between the hybridized plasmonic (HP) and hybridized waveguide (HW) modes, we define the overlap parameter as
  
\begin{align}
  & overlap=\operatorname{Re}\left[ \frac{\left( \int{{{{\vec{E}}}_{HP}}\times \vec{H}_{HW}^{*}.\,d\vec{S}} \right)\left( \int{{{{\vec{E}}}_{HW}}\times \vec{H}_{HP}^{*}.\,d\vec{S}} \right)}{\int{{{{\vec{E}}}_{HP}}\times \vec{H}_{HP}^{*}.\,d\vec{S}}} \right]\nonumber\\ 
 & \quad \quad \quad \quad \quad .\frac{1}{\operatorname{Re}\left( \int{{{{\vec{E}}}_{HW}}\times \vec{H}_{HW}^{*}.\,d\vec{S}} \right)} 
\end{align}
Figure 3(b) displays the variation of this overlap as a function of the layer width and the height of the top Si layer. The maximum overlap value of 0.977 is achieved at a layer width of 600 nm and a top Si height of 310 nm, corresponding to the EP degeneracy condition. 
\begin{figure*}
\centering
{\includegraphics[width=.8\textwidth]{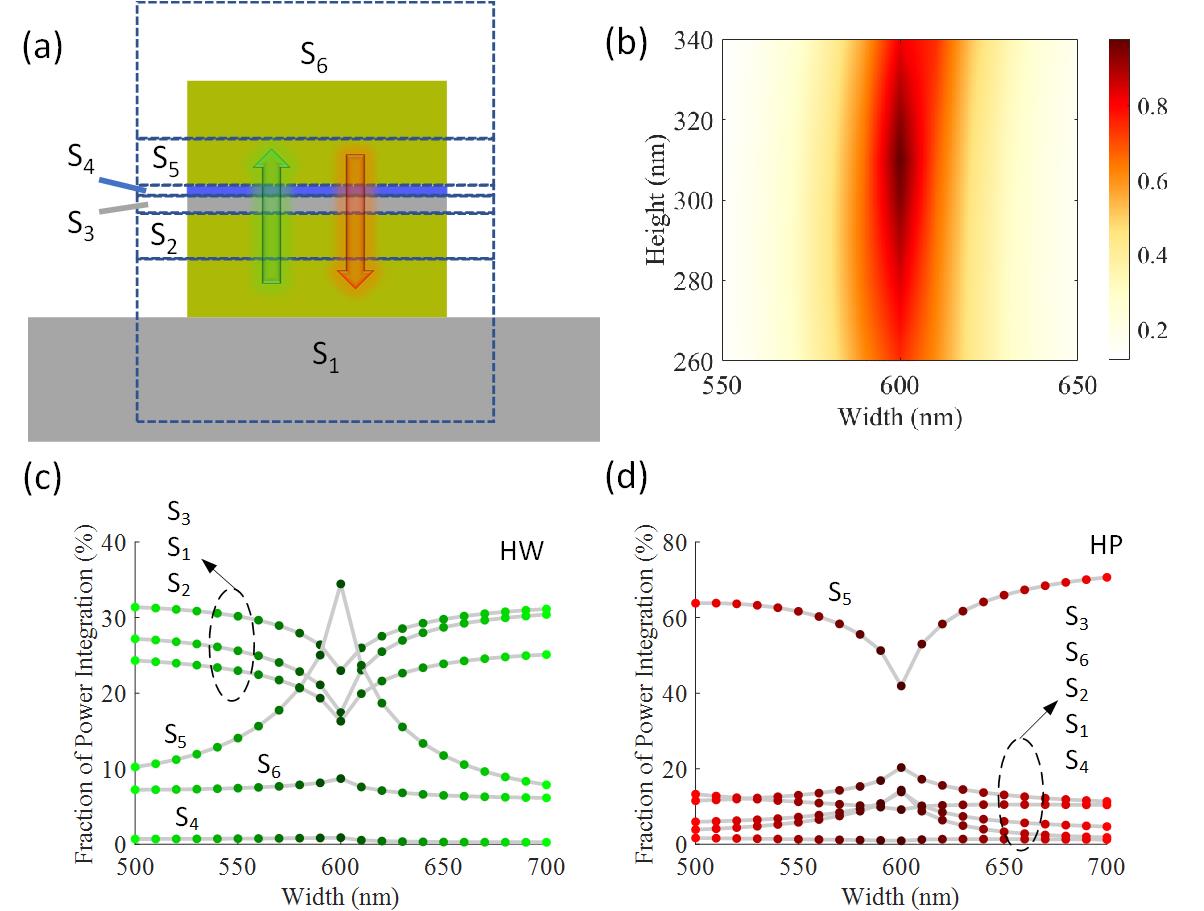}}
\caption{To elucidate the dynamic power transfer induced by variations in the geometrical dimensions, the hybrid structure is partitioned into discrete regions (S$_1$–S$_6$) for detailed modal power fraction analysis. In the nearly decoupled regime, the hybrid waveguide (HW) mode primarily confines its energy below the metallic layer, whereas the hybrid plasmonic (HP) mode concentrates most of its power above the metal. As the system parameters are tuned toward the exceptional point (EP), the HW mode progressively shifts its power upward, and the HP mode correspondingly transfers power downward, ultimately achieving a balanced condition at the EP where the two modes degenerate with an overlap approaching unity. (a) The partitioning strategy used for the power calculations. (b) The overlap between the HP and HW modes, with a maximum observed at a structure width of 600 nm and height of 310 nm, corresponding to EP degeneracy. (c) and (d) The power transfer dynamics across regions S$_1$–S$_6$ for the HW and HP modes, respectively, highlighting the abrupt changes in modal power distribution as the geometry is varied.}
\end{figure*}
\section{Results}
To comprehensively analyze the characteristics of exceptional point (EP) degeneracy, we performed full-wave 3D simulations, capturing both the field distribution and transmission spectra while varying the input wavelength and tuning the material properties of a functional layer. Specifically, we utilized a low-loss phase change material (PCM) to dynamically control the system’s effective parameters, enabling precise tuning toward the EP regime.  Notably, the transmission spectra were evaluated over a 900 nm long structure, demonstrating substantial amplitude modulation within an ultra-compact footprint. 

Figures 4(a) and 4(b) illustrate the real and imaginary components of the effective refractive index as functions of the input wavelength, confirming their degeneracy at \(\lambda = 1550\) nm. The degeneracy of \(\operatorname{Re}(n_{\text{eff}})\) at this wavelength signifies the modal coalescence characteristic of EPs, while the merging of \(\operatorname{Im}(n_{\text{eff}})\) implies a balanced dissipation mechanism. Furthermore, the transmission spectra, shown in Figure 4(c), closely follow the trend of \(\operatorname{Im}(n_{\text{eff}})\), reinforcing the role of loss modulation in governing the system’s spectral response. The observed dissipation-driven transmission behavior highlights the nontrivial interplay between waveguide and plasmonic modes, particularly in proximity to the EP.  

To further validate the EP characteristics, in Figs. 4(d) and 4(e), we examine the electric field distributions in both longitudinal and transverse cross-sections, respectively. The longitudinal field profile, extracted at the midplane of the silica layer, reveals nearly indistinguishable distributions at EP, in contrast to the distinctly separated modal profiles observed in the weakly coupled regime. Additionally, the transverse field components exhibit nearly identical spatial distributions at the degenerate point, signifying strong modal hybridization. This distinct transition in the field distribution confirms the fundamental restructuring of eigenmodes, marking the onset of non-Hermitian degeneracy and the breakdown of conventional mode orthogonality.  
\begin{figure*}
\centering
{\includegraphics[width=.8\textwidth]{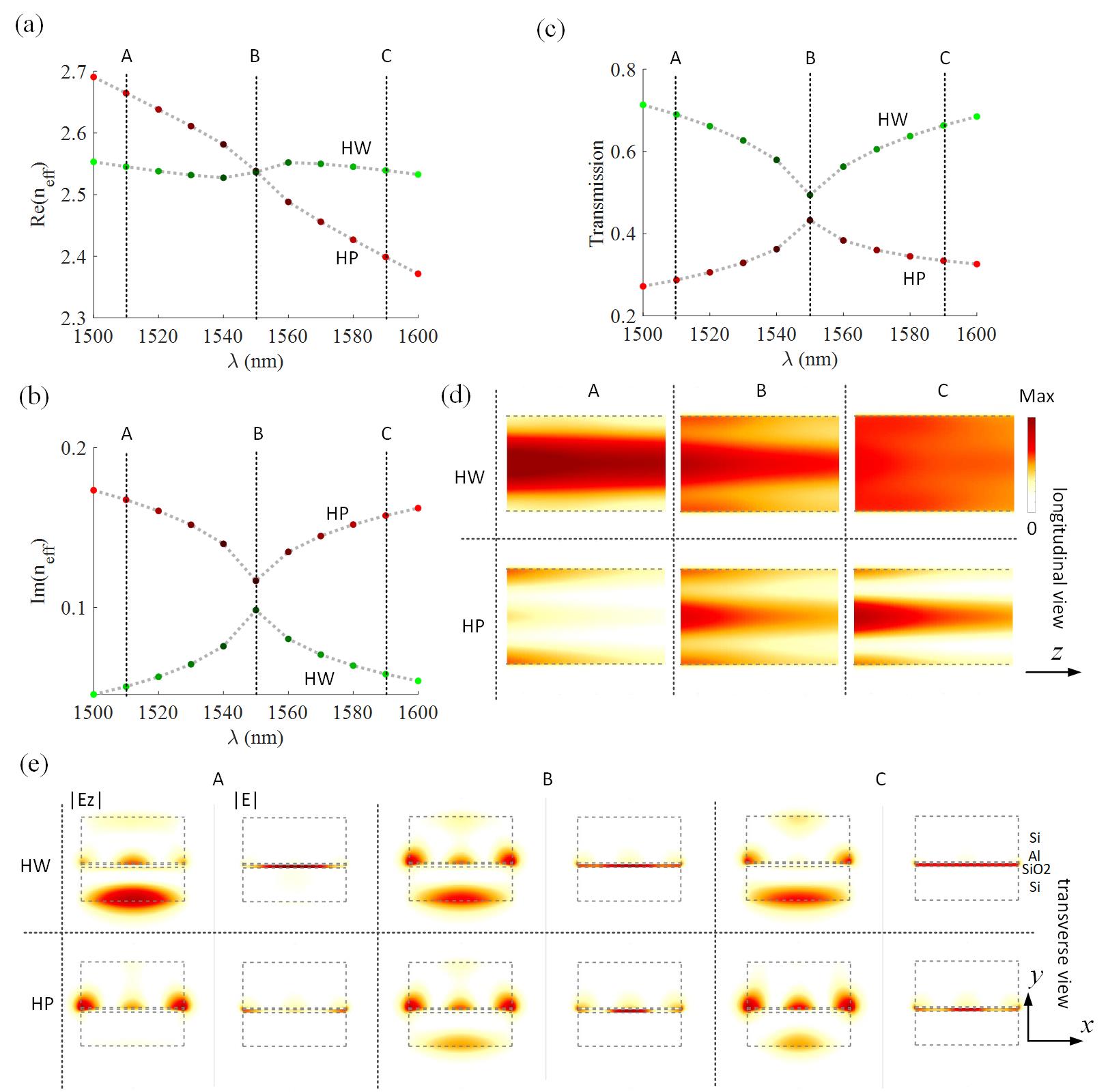}}
\caption{Frequency-dependent tunability of the hybrid structure is demonstrated through the effective index and transmission spectra as functions of the input wavelength over a propagation distance of 900 nm. (a) and (b) The real and imaginary parts of the effective refractive index, respectively, with an exceptional point (EP) established at 1550 nm. (c) The transmission spectrum, which mirrors the behavior of the imaginary index. (d)-(e) The electric field distributions in both the (d) longitudinal and (e) transverse views reveal that at the EP the field profiles of the two modes become nearly identical. Notably, the hybrid waveguide mode exhibits maximum loss at the EP, with its loss decreasing abruptly when moving away from this point, whereas the hybrid plasmonic mode shows a minimum loss at the EP that increases sharply outside the EP.}
\end{figure*}

In Fig. 5, we investigate the electrically tunable properties of a hybrid plasmonic waveguide structure by integrating a 40‐nm phase-change material (PCM) layer—either Sb$_2$S$_3$ or Sb$_2$Se$_3$—between an aluminum layer and a top silicon layer. Sb$_2$S$_3$ and Sb$_2$Se$_3$ serve as ultralow loss phase-change materials (PCMs) for photonic integrated circuits (For a detailed analysis of how varying the PCM thickness influences the results, please see Supplemental Material, Section I) \cite{suppmat2025}. At a wavelength of 1550 nm, Sb$_2$S$_3$ exhibits refractive indices of approximately 2.7 in its amorphous phase and 3.3 in its crystalline phase, while Sb$_2$Se$_3$ shows values around 3.3 and 4, respectively \cite{del}. Recent advancements further highlight the promise of Sb$_2$Se$_3$, demonstrating switching energies of 9.25 nJ for amorphization and 1280 nJ for crystallization, alongside excellent endurance characteristics \cite{fang2022ultra, alam2024fast}. These findings underscore the potential of these PCMs for enabling highly efficient and reliable photonic devices.

The PCM serves as a tunable element due to its low loss and a refractive index that, in its optimal phase (crystallized for Sb$_2$S$_3$ and amorphous for Sb$_2$Se$_3$), closely matches that of silicon. In the first configuration, we employ Sb$_2$S$_3$ and adjust the geometrical parameters so that the device reaches an exceptional point (EP) when the material is crystallized; switching Sb$_2$S$_3$ to its amorphous state shifts the operation away from the EP. Conversely, in the second configuration, Sb$_2$Se$_3$ is used, with the structure optimized to satisfy EP conditions in its amorphous phase, while the crystallized state results in an operating point outside the EP regime.

To further elucidate the effect of the PCM on the modal properties, we analyzed the effective refractive indices as functions of the PCM refractive index. Figures 5(a) and 5(b) display the real and imaginary parts, respectively, for Sb$_2$S$_3$, confirming that the EP is achieved in the crystallized state and lost upon amorphization. Similarly, Figures 5(c) and 5(d) present the effective index components for Sb$_2$Se$_3$, where the EP occurs in the amorphous phase and is absent in the crystallized state. By mapping the imaginary part of effective index to propagation loss, we can quantitatively assess the system’s dissipative behavior. For instance, for Sb\(_2\)S\(_3\), the loss is approximately 3.14 dB/$\mu$m at the EP, which decreases sharply to 0.63 dB/$\mu$m when operating away from the EP. Similarly, Sb\(_2\)Se\(_3\) exhibits a propagation loss of around 3.00 dB/$\mu$m at the EP, reducing to 0.89 dB/$\mu$m outside the EP regime. These variations in loss are directly reflected in the transmission characteristics: the transmission coefficient (T) shifts from approximately 0.53 (or 0.54) at the EP to about 0.88 (or 0.83) out of the EP. This mapping of the imaginary part of the effective index to propagation loss—and subsequently to transmission—provides a clear and quantitative insight into the non-Hermitian dynamics governing the system, paving the way for the design of ultracompact modulators and high-sensitivity photonic devices. Figure 5(e) schematically illustrates the hybrid plasmonic waveguide, emphasizing the integration of the low‐loss PCM. Finally, Fig. 5(f) presents a longitudinal cross-sectional view of the electric field distribution within the central region of the silica layer over a 900‐nm propagation length. The longitudinal field profile, extracted at the midplane of the silica layer, reveals nearly indistinguishable distributions at the EP, in contrast to the distinctly separated modal profiles observed in the weakly coupled regime. 

\begin{figure*}
\centering
{\includegraphics[width=.8\textwidth]{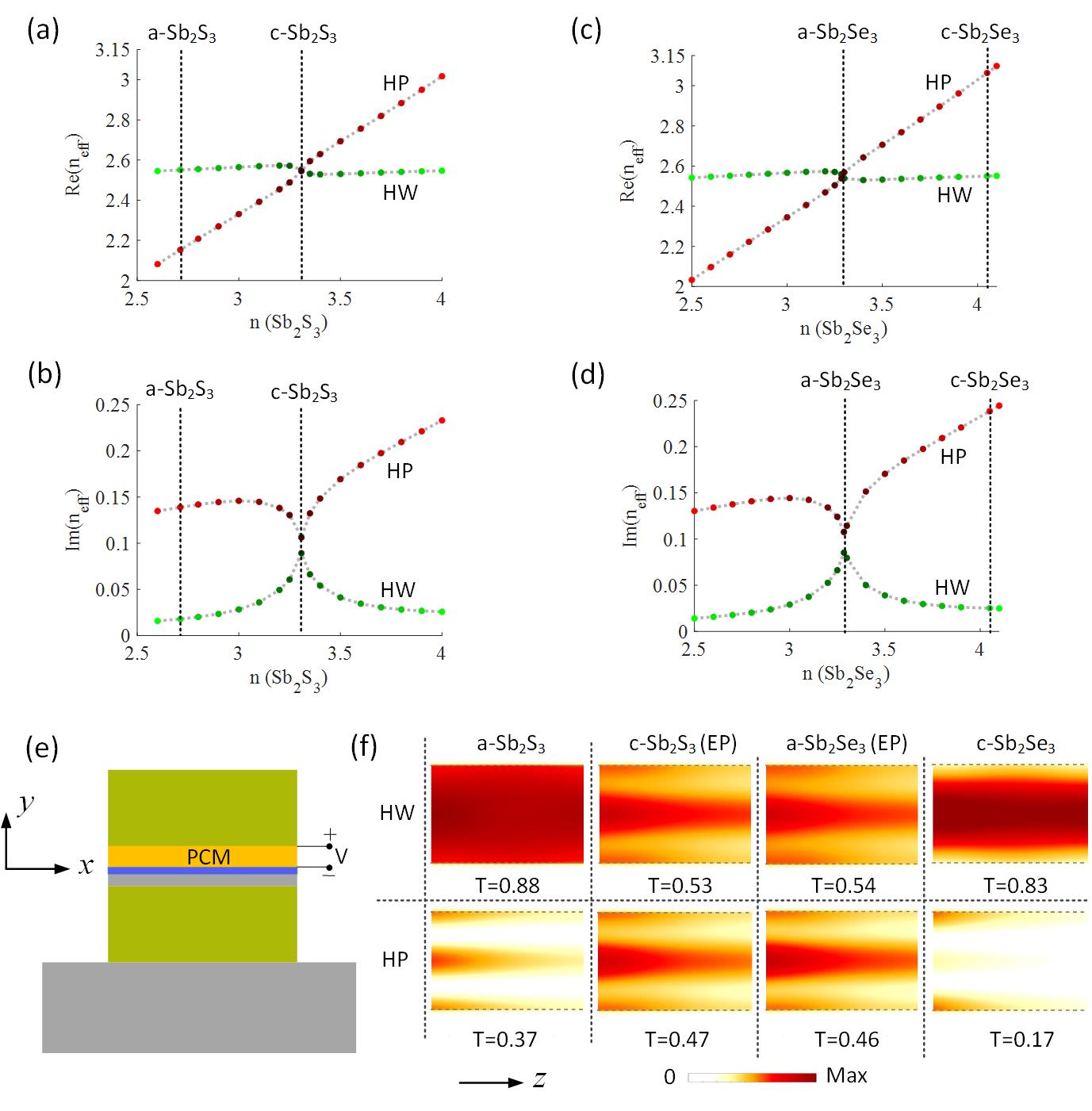}}
\caption{A 40‐nm phase-change material (PCM) layer—either Sb$_2$S$_3$ or Sb$_2$Se$_3$—is interposed between the aluminum layer and top silicon layer to serve as a tunable element. Notably, the low‐loss PCM exhibits a refractive index nearly identical to that of silicon in its favorable phase (crystallized for Sb$_2$S$_3$ and amorphous for Sb$_2$Se$_3$). In the first configuration, the device incorporates Sb$_2$S$_3$ with the geometrical parameters adjusted to reach an exceptional point (EP) when the material is crystallized; switching to the amorphous state shifts the operation away from the EP. Conversely, in the second configuration, Sb$_2$Se$_3$ is employed with the structure optimized for EP conditions in its amorphous phase, while its crystallized phase moves the system out of the EP regime. (a) and (b) display the real and imaginary parts of the effective refractive index as functions of the Sb$_2$S$_3$ refractive index, confirming that the EP is achieved in the crystallized state and lost upon amorphization. (c) and (d) The effective index components versus the Sb$_2$Se$_3$ refractive index, where the EP occurs in the amorphous phase and is absent in the crystallized state. (e) The hybrid plasmonic waveguide schematic, highlighting the integration of the low‐loss PCM. (f) The longitudinal view of electric field distribution within the central region of the silica layer over a 900‐nm propagation length, evidencing the transmission characteristics and strong amplitude modulation.}
\end{figure*}
\section{discussion}
Despite significant advances in electro-optical modulation within silicon (Si) integrated photonics—especially through the incorporation of advanced functional materials—current techniques remain constrained by limitations in dimensionality and tunability. To break through these barriers, new physical concepts and device architectures are necessary. In our study, we leverage non-Hermitian physics, specifically the phenomenon of exceptional points (EPs), to explore a fundamentally different approach to modulator design. Exceptional points, which represent degeneracies in non-Hermitian systems, offer unique opportunities for mode control and sensitivity enhancement in ultracompact devices.

Our device is based on a layered structure featuring a thin film of SiO$_2$/Al (with respective thicknesses of 20 nm and 10 nm) integrated between Si layers. This configuration supports multiple orders of hybrid plasmonic waveguide modes that exhibit EP characteristics. The interplay between plasmonic and dielectric waveguide modes facilitates both strong and weak coupling regimes, thereby enabling the formation of waveguide-plasmon polariton quasiparticles. By carefully tailoring the spatial parameters of the structure, we achieve conditions where these modes possess identical effective indices and field distributions, marking the benchmark of EP degeneracy.

To demonstrate electrical tunability, we incorporated a 40-nm-thick layer of low-loss phase-change materials (PCMs), specifically Sb$_2$S$_3$ and Sb$_2$Se$_3$, between the metal layer and the top Si. At a wavelength of 1.55 $\mu$m, the hybrid waveguide mode exhibits a sharp transition in propagation loss. For Sb$_2$S$_3$, the loss varies from approximately 3.14 dB/$\mu$m at the EP to 0.63 dB/$\mu$m out of the EP. For Sb$_2$Se$_3$, the corresponding variation is from 3.00 dB/$\mu$m at the EP to 0.89 dB/$\mu$m out of the EP. These abrupt changes in loss yield a transmission coefficient (T) that shifts from T$\approx$0.53 (or 0.54) at the EP to T$\approx$0.88 (or 0.83) out of the EP, all within a propagation length of just 0.9 $\mu$m. Such sensitivity highlights the effectiveness of the EP in modulating the optical response.

For further characterization, beyond eigenmode dispersion and idealized mode‐launching simulations, we excite the structure using a single silicon access waveguide (see Section II of the Supplemental Material) \cite{suppmat2025}. Although this approach launches a superposition of the structure’s eigenmodes, the hallmark of the exceptional point remains clearly discernible in the resulting propagation spectrum. Achieving more efficient and mode‐selective coupling—from the strip‐waveguide fundamental mode into the slot-plasmon mode—will require tailored input‐waveguide designs (for example, a slit or taper) \cite{saynatjoki2011low,wang2009ultracompact,yang2010efficient,butt2020ultrashort,zhu2016high,mere2018efficient,han2016strip,nair2024compact,li2013efficient}, which lies beyond the scope of the present study and will be addressed in future work.

One promising avenue is to incorporate gain media or utilize complex-frequency excitation to induce virtual gain. In such schemes, the inherent loss at the EP could be precisely counterbalanced, effectively reaching a zero net-loss condition. This compensation not only sharpens the EP response but also enhances the overall responsivity of the device—a critical factor for applications in on-chip optical modulation and sensing.
The exploration of EPs in plasmonic structures opens a new frontier in fundamental light-matter interaction research at the nanoscale. Beyond fundamental science, the proposed structure paves the way for a host of industrial applications in ultracompact photonic devices, optical communication, sensing and metrology.

\section{Conclusion}
Our work demonstrates that by harnessing the unique properties of exceptional points within a carefully engineered Si integrated modulator, it is possible to overcome current limitations in dimensionality and tunability. The combination of hybrid plasmonic waveguide modes and phase-change material-induced electrical tuning not only provides superior modulation characteristics but also opens up exciting avenues for both fundamental research and practical device applications. Future explorations—especially those involving gain integration and alternative tunable materials—are expected to further enhance the performance and versatility of integrated photonic devices.

\bibliography{bibliography}% Produces the bibliography via BibTeX.

\end{document}